\documentclass[]{aastex62}
\usepackage{color}
\usepackage{graphicx}
\usepackage{longtable}
\usepackage{multirow}
\usepackage{url}
\usepackage{subfigure}
\usepackage{amsmath}
\usepackage{lipsum}
\usepackage{txfonts}
\usepackage{natbib}
\usepackage{ucs}
\shorttitle{FRB 190520B embedded in a composite of MWN and SNR}
\shortauthors{Zhao \& Wang}

\received{Nov 25, 2021}
\revised{Nov 30, 2021}
\accepted{Dec 2, 2021}

\begin{document}
\title{FRB 190520B embedded in a magnetar wind nebula and supernova remnant: luminous persistent radio source, decreasing dispersion measure and large rotation measure}

\author[0000-0002-2171-9861]{Z. Y. Zhao}
\affiliation{School of Astronomy and Space Science, Nanjing University, Nanjing 210093, China}

\author[0000-0003-4157-7714]{F. Y. Wang}
\affiliation{School of Astronomy and Space Science, Nanjing University, Nanjing 210093, China}
\affiliation{Key Laboratory of Modern Astronomy and Astrophysics (Nanjing University), Ministry of Education, Nanjing 210093, China}

\correspondingauthor{F. Y. Wang}
\email{fayinwang@nju.edu.cn}

\begin{abstract}
Recently, FRB 190520B with the largest extragalactic dispersion measure (DM), was discovered by the Five-hundred-meter Aperture Spherical radio Telescope (FAST). The DM excess over the intergalactic medium and Galactic contributions is estimated as $\sim 900$ pc cm$^{-3}$, which is nearly ten times higher than other fast radio bursts (FRBs) host galaxies. The DM decreases with the rate $\sim0.1$ pc cm$^{-3}$ per day. It is the second FRB associated with a compact persistent radio source (PRS). The rotation measure (RM) is found to be larger than $1.8 \times 10^{5} \mathrm{rad} ~\mathrm{m}^{-2}$. In this letter, we argue that FRB 190520B is powered by a young magentar formed by core-collapse of massive stars, embedded in a composite of magnetar wind nebula (MWN) and supernova remnant (SNR). The energy injection of the magnetar drives the MWN and SN ejecta to evolve together, and the PRS is generated by the synchrotron radiation of the MWN. The magnetar has the interior magnetic field $B_{\text{int}}\sim (2-4)\times 10^{16}$ G and the age $t_{\text{age}}\sim 14-22$ yr.
The dense SN ejecta and the shocked shell contribute a large fraction of the observed DM and RM. Our model can naturally explain the luminous PRS, decreasing DM and extreme RM of FRB 190520B simultaneously. 
\end{abstract}


\section{Introduction}
Fast radio bursts (FRBs) are mysterious radio transients with millisecond-duration \citep{Lorimer2007}, whose physical origins are still unknown though they were first reported more than a decade ago \citep{Katz2018,Cordes2019,ZB2020,Xiao2021,Petroff2021}. The large dispersion measures (DMs) of them well above the contribution from the Milky Way imply they may originate at cosmological distances. Some FRBs show repeating bursts and other seem to be one-off events. Many models have been proposed to interpret the origins of FRBs (see \citealt{Platts2019} for a recent review). Among those models, the ones relational to magnetars are promising because of the detection of FRB 200428 from a Galactic magnetar \citep{Bochenek2020,CHIME2020}.

Recently, the repeating FRB 190520B with the largest extragalactic DM till now was discovered by the Five-hundred-meter Aperture Spherical radio Telescope (FAST) \citep{Niu2021}. It locates in a dwarf galaxy with a high star formation rate, and it is associated with a compact luminous ($\nu L_{\nu} \sim 10^{39}$ erg s$^{-1}$) persistent radio source (PRS), which is too luminous to come from the star-formation activity of the host galaxy \citep{Law2021}. 
From the observations of Karl G. Jansky Very Large Array (VLA), the power-law spectrum index of the compact PRS has been found to be $-0.41\pm0.04$. This is the second PRS associated with FRBs after FRB 121102 \citep{Chatterjee2017}. The similarity between the two PRSs indicates that they have similar physical origin. Interestingly, a similar PRS is found associated with Type I superluminous supernova (SLSN) PTF10hgi, but is less luminous ($\nu L_{\nu} \sim 10^{38}$ erg s$^{-1}$) than that of FRBs \citep{Eftekhari2019}. From the comparative study between the wide-band spectrum of PTF10hgi and FRB 121102 \citep{Mondal2020}, they found the PRS is most probably originating from a pulsar/magnetar wind nebula (PWN/MWN). It has been found that magnetar Swift J1834.9-0846 shows a surrounding wind nebula \citep{Younes2016}.
In this work, we focus on the PRSs associated with FRBs.

When a pulsar-driven relativistic wind interacts with the surrounding medium, the luminous PWN generates. For rapidly rotating pulsars, the rotational energy is the main reservoir for powering the wind nebula, which has been well-studied for Galactic PWNe \citep{Tanaka2010}. Some FRBs' energy injection \citep{Li2020} and rotational energy injection \citep{Kashiyama2017,Dai2017,Yang2019,Wang2020} models have been proposed to explain the PRS associated with FRB 121102. However, for a decades-old magnetar, the rotational energy is less significant than the interior magnetic energy. The case of the magnetic energy injection was proposed, and it is successfully explaining the PRS's luminosity and large rotation measure (RM, \citealt{Michilli2018,Hilmarsson2021}) of FRB 121102 \citep{Margalit2018}. 

From the estimation of \cite{Niu2021}, the DM of host galaxy is DM$_{\text{host}}\sim 900$ pc cm$^{-3}$, which is nearly ten times higher than other FRBs' host galaxies. However, using the state-of-the-art IllustrisTNG simulation, \cite{Zhang2020} showed that the DM contributed by FRB 190520B-like host galaxies at $z\sim0.2$ is $\sim50-250$ pc cm$^{-3}$. Different from the increasing DM of FRB 121102 \citep{Hessels2019,Josephy2019,Oostrum2020,Li2021} and the nearly unchangeable DM of FRB 180916 \citep{2020Natur.582..351C,Pastor-Marazuela2021,Nimmo2021}, the DM of FRB 190520B decreases with the rate $-0.09 \pm 0.02$ pc cm$^{-3}$ day$^{-1}$ \citep{Niu2021}. Under the assumption of 100\% linearly polarized intrinsically, the low limit of  RM is $>1.8 \times 10^{5}~ \mathrm{rad} ~\mathrm{m}^{-2}$ \citep{Niu2021}, which is larger than that of FRB 121102. The large DM and RM together with decreasing DM may come from the expanding shocked shell of supernova remnant (SNR, see \citealt{Yang2017,Piro2018,Zhao2021,Katz2021}). \cite{Katz2021} rejected the possibility that the large host DM of FRB 190520B is contributed by interstellar cloud, and proposed that the excess of host DM attributes to a young SNR. It has also been claimed that the luminous PRS correlates with the large RM, if RM mostly arises from the persistent emission region \citep{Yang2020}.

In this letter, we propose that the magnetar associated with FRB 190520B is embedded in the composite of MWN and SNR. The magnetar is formed by core-collapse of massive star. Due to the energy injection of the young magnetar, the wind nebula and the SN ejecta will evolve together. The observed PRS is produced by the synchrotron radiation of the nebula. Our numerical calculations are based on the spectral evolution model of Galactic PWNe \citep{Tanaka2010,Tanaka2013} with magnetic-energy injection \citep{Margalit2018}. 
The dense SNR ejecta and the shocked shell contribute considerable DM and RM. With the expansion of SNR, the DM will decrease rapidly, similar to the observed trend. Our model can simultaneously explain the luminous PRS, decreasing DM and extreme RM of FRB 190520B. 

This letter is organized as follows. In Section \ref{PRS}, the synchrotron spectral evolution model from MWN is shown. We present our numerical results of the PRS energy spectrum in Section \ref{result}. The long-term DM evolution model to explain that of FRB 190520B is shown in Section \ref{DM}. Finally, summary is given in Section \ref{conclusion}.

\section{The compact persistent radio source}\label{PRS}
The compact PRS associated with FRBs is from the synchrotron radiation from the MWN powered by the young magnetar in our model. Rotational or magnetic energy together with the particle is injected into the nebula, and the electron will undergo radiation or adiabatic cooling. In this section, we will introduce the cases of rotational and magnetic energy injection, and give the explanation of the radio spectra of PRSs associated with FRB 190520B and FRB 121102.
\subsection{Energy injection}
The case of rotational energy injection is well studied for wide-band spectrum of the Crab Nebula \citep{Tanaka2010}. The spin-down luminosity can be estimated as \citep{Dai1998,ZHang2001,Murase2015}
\begin{equation}
\begin{aligned}
    L_{\mathrm{sd}}=& L_{\mathrm{sd}, i}\left(1+\frac{t}{t_{\mathrm{em}}}\right)^{-2} 
    & \simeq \begin{cases}8.6 \times 10^{45} \mathrm{erg} \mathrm{s}^{-1} P_{i,-2.5}^{-4} B_{\mathrm{dip}, 14}^{2} & \left(t \leqslant t_{\mathrm{em}}\right) \\
    8.9 \times 10^{42} \mathrm{erg} \mathrm{s}^{-1} B_{\mathrm{dip}, 14}^{-2} t_{7}^{-2} & \left(t>t_{\mathrm{em}}\right)\end{cases},
\end{aligned}
\end{equation}
where $t_{\mathrm{em}}\simeq 3.2 \times 10^{5} B_{\mathrm{dip}, 14}^{-2} P_{i,-2.5}^{2}\mathrm{~s} $ is the characteristic spin-down timescale, $B_{\mathrm{dip}}$ is the dipole magnetic and $P_{i}$ is the initial spin period. 
The model of magnetic energy injection has been proposed to explain the exceptionally high RM and PRS associated with FRB 121102 \citep{Margalit2018}. The interior magnetic energy \citep{Katz1982}
\begin{equation}
    \mathcal{E_{\text{B,int}}} \simeq B_{\text{int}}^{2} R_{\text{ns}}^{3} / 6 \approx 3 \times 10^{49} B_{\text{int},16}^{2} \text{~erg},
\end{equation}
is another ideal reservoir for PRS, where $B_{\text{int}}$ is the interior magnetic field and $R_{\text{ns}}=12$ km is the neutron star radius. The magnetic-energy-injection luminosity can be written as \citep{Margalit2018}
\begin{equation}
    L_{\mathrm{m}}=(\alpha-1) \frac{\mathcal{E_{\text{B,int}}} }{t_{0}}\left(\frac{t}{t_{0}}\right)^{-\alpha},
\end{equation}
where $t_0$ is the onset of energy injection and $\alpha>1$ is the power-law index. 

The injected electron-positron pairs will be accelerated to relativistic energy by the termination shock before entering the nebula. Similar to Galactic PWNe \citep{Tanaka2010,Tanaka2013}, the injection particles spectrum is described as a broken power-law form
\begin{equation}
    Q_{\mathrm{inj}}(\gamma, t)= \begin{cases}Q_{0}(t)\left(\gamma / \gamma_{\mathrm{b}}\right)^{-p_{1}} &  \quad \gamma_{\min} \leqslant \gamma \leqslant \gamma_{\mathrm{b}} \\ Q_{0}(t)\left(\gamma / \gamma_{\mathrm{b}}\right)^{-p_{2}} &  \quad \gamma_{\mathrm{b}} \leqslant \gamma \leqslant \gamma_{\max }\end{cases},
\end{equation}
where $Q_{0}(t)$ is the normalization factor, $\gamma_{\min}$, $\gamma_{b}$ and $\gamma_{\max}$ is the minimum, break, and maximum Lorentz factors. $p_1$ and $p_2$ are the injection spectrum indices for low and high energy particles, respectively. The normalization factor is determined by
\begin{equation}
     \int_{\gamma_{\min}}^{\gamma_{\max }} Q_{\mathrm{inj}}(\gamma, t) \gamma m_{e} c^{2} d \gamma = \epsilon_{\mathrm{e}} L(t),
\end{equation}
where $\epsilon_{\mathrm{e}}$ is the electron energy fraction and $L(t)$ is the spin-down or magnetic energy injection luminosity. 

\subsection{Dynamics and the nebular magnetic fields evolution}
\begin{figure}
    \centering
    \includegraphics[width = 1\textwidth]{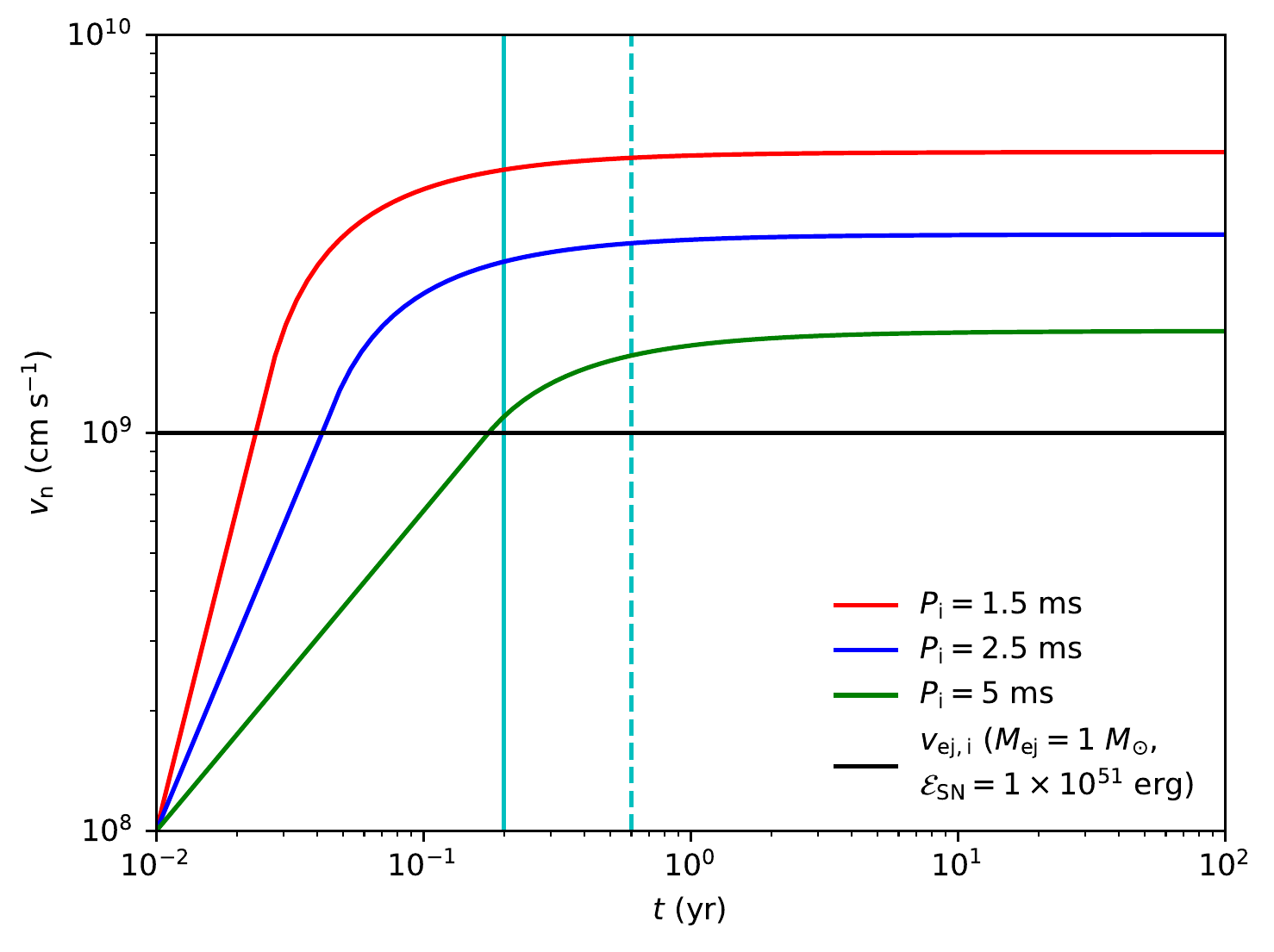}
    \caption{The nebula velocity derived from Equations (\ref{Rn1}) and (\ref{Rn2}) for $B_{\text{int}}\sim10^{16}$ G, $M_{\mathrm{ej}}=1~M_{\odot}$, and $\mathcal{E}_{\mathrm{SN}}=1 \times 10^{51}$ erg. The red, blue and green solid lines represent the case of $P_{\text{i}}=1.5$ ms, $P_{\text{i}}=2.5$ ms and $P_{\text{i}}=5$ ms, respectively. The initial ejecta velocity is shown in black lines. The onset of the magnetic energy injection $t_0=0.2$ yr and 0.6 yr is shown in cyan solid and dashed line, respectively. We can see that ejecta is accelerated significantly in a relatively short time ($\sim 10^{-2}-10^{-1}$ yr) before the magnetic flux begins to leak out. The finial ejecta velocity is up to $\sim 20,000-60,000 \mathrm{~km} \mathrm{~s}^{-1}$, which is consistent with the observations of SN Ib/Ic.}
    \label{vn}
\end{figure}

The inner density profile of the ejecta can be described as a smooth or flat power-law \citep{Chevalier1989,Kasen2010}
\begin{equation}
    \rho_{\mathrm{ej}}=\frac{(3-\delta) M_{\mathrm{ej}}}{4 \pi R_{\mathrm{ej}}^{3}}\left(\frac{R}{R_{\mathrm{ej}}}\right)^{-\delta},
\end{equation}
where $\delta=0-1$ is widely used, and we take $\delta=1$ in this work. The ejecta will expand freely until the Sedov–Taylor phase without the energy injection. The initial velocity is $v_{\mathrm{ej}, \mathrm{i}}\sim 10,000 \mathrm{~km} \mathrm{~s}^{-1}\left(\mathcal{E}_{\mathrm{SN}} / 10^{51} \mathrm{erg}\right)^{1 / 2}\left(M_{\mathrm{ej}} / M_{\odot}\right)^{-1 / 2} \mathrm{~cm} \mathrm{~s}^{-1}$. When a newborn millisecond magnetar exists, the nebula and ejecta radius will evolve together because the injected energy will significantly accelerate the ejecta via magnetized wind. For $R_{\text{n}}<R_{\text{ej}}$, the nebula radius $R_{\text{n}}$ is given by \citep{Metzger2014} 
\begin{equation}\label{Rn1}
    \frac{d R_{n}}{d t}=\sqrt{\frac{7}{6(3-\delta)} \frac{\mathcal{E}_{\mathrm{tot}}}{M_{\mathrm{ej}}}\left(\frac{R_{n}}{R_{\mathrm{ej}}}\right)^{3-\delta}}+\frac{R_{n}}{t},
\end{equation}
where $\mathcal{E}_{\mathrm{tot}}$ is the total injection energy. If $R_{\text{n}}>R_{\text{ej}}$, the nebula and ejecta will move together
\begin{equation}\label{Rn2}
    \frac{\mathrm{d} R_{\mathrm{ej}}}{\mathrm{d} t}=\frac{\mathrm{d} R_{\mathrm{n}}}{\mathrm{d} t}=v_{\mathrm{ej,f}},
\end{equation}
where $v_{\mathrm{ej,f}}=\sqrt{2(\mathcal{E}_{\mathrm{tot}}+\mathcal{E}_{\mathrm{SN}})/M_{\mathrm{ej}}}$ is the finial accelerated velocity. For $t<t_0$, the rotational energy injection dominates, and the injection energy is $\mathcal{E}_{\mathrm{tot}}=\mathcal{E}_{\mathrm{rot}}=\int_{0}^{t} L_{\mathrm{sd}}(t) \mathrm{d} t$. For $t>t_0$, the interior magnetic energy starts to leak out into the nebula and the total injection energy $\mathcal{E}_{\mathrm{tot}}=\mathcal{E}_{\mathrm{rot}}+\mathcal{E}_{\mathrm{mag}}=\int_{0}^{t} (L_{\mathrm{sd}}(t) +L_{\mathrm{m}}(t)) \mathrm{d} t$. The solution of Equations (\ref{Rn1}) and (\ref{Rn2}) for the example case $B_{\text{int}}\sim10^{16}$ G, $M_{\mathrm{ej}}=1~M_{\odot}$, and $\mathcal{E}_{\mathrm{SN}}=1 \times 10^{51}$ erg is shown in Figure \ref{vn}. The red, blue and green solid lines represent the case of $P_{\text{i}}=1.5$ ms, $P_{\text{i}}=2.5$ ms and $P_{\text{i}}=5$ ms, respectively. The initial ejecta velocity is shown as black lines. The onset of the magnetic energy injection $t_0=0.2$ yr and 0.6 yr from the benchmark model of \cite{Margalit2018} is shown in cyan solid and dashed lines, respectively. We can see that ejecta is accelerated significantly in a short time ($\sim 10^{-2}-10^{-1}$ yr) before the magnetic flux begins to leak out. The finial ejecta velocity is up to $\sim 20,000-60,000 \mathrm{~km} \mathrm{~s}^{-1}$, which is well consistent with the observations of SN Ib/Ic \citep{Kawabata2002,Rho2021}.

The evolution of nebular magnetic fields is $B_{\mathrm{n}}=\sqrt{(6 \mathcal{E_{\text{B}}} / R_{\mathrm{n}}^{3})}$, where $\mathcal{E_{\text{B}}}$ is the magnetic energy in nebular and $R_{\mathrm{n}}$ is the nebula radius. The magnetic energy in nebular is given by \citep{Murase2021}
\begin{equation}\label{Bn}
    \frac{d \mathcal{E_{\text{B}}}}{dt}=\epsilon_{\mathrm{B}}L(t)-c_{\mathrm{B}}\frac{\dot{R}_{\mathrm{n}}}{R_{\mathrm{n}}} \mathcal{E_{\text{B}}},
\end{equation}
where $\epsilon_{\mathrm{B}}$ is the magnetic energy fraction. In this work, we do not consider the magnetic energy loss caused by adiabatic expansion, which has been used in Galactic PWNe \citep{Tanaka2010,Tanaka2013} and high-energy emission of pulsar-powered PWNe \citep{Murase2015,Murase2016}. The limit $c_{\mathrm{B}}\to0$ is a good approximation for a young source engine.

\subsection{The Evolution of Particle Distribution}
The evolution of the electron number density distribution $n_{e,\gamma}$ is given by the continuity equation in energy space 
\begin{equation}\label{ne}
    \frac{\partial}{\partial t} n_{e,\gamma}+\frac{\partial}{\partial \gamma}\left(\dot{\gamma} n_{e,\gamma}\right)=\dot{Q}_{e,\gamma}, 
\end{equation}
where $\dot{Q}_{e,\gamma}$ is the injection electron number density. The electron cooling process $\dot{\gamma}$ includes the synchrotron radiation, synchrotron self-Compton (SSC) and adiabatic expansion
\begin{equation}
    \dot{\gamma}(\gamma, t)=\dot{\gamma}_{\mathrm{syn}}(\gamma, t)+\dot{\gamma}_{\mathrm{SSC}}(\gamma,t)+\dot{\gamma}_{\mathrm{ad}}(\gamma, t).
\end{equation}
The energy loss of synchrotron radiation is given by \citep{Rybicki1979}
\begin{equation}
    \dot{\gamma}_{\mathrm{syn}}(\gamma, t)=-\frac{4}{3} \frac{\sigma_{\mathrm{T}}\gamma^{2}}{m_{e} c}  U_{\mathrm{B}}(t), 
\end{equation}
where $U_{\mathrm{B}}=B_{\text{n}}^2/8 \pi$ is the energy density of magnetic field. The energy loss caused by SSC is \citep{Blumenthal1970}
\begin{equation}
\begin{aligned}
    \dot{\gamma}_{\mathrm{SSC}}(\gamma,t)=&-\frac{3}{4}  \frac{\sigma_{\mathrm{T}} h}{m_{e} c}  \frac{1}{\gamma^{2}} \int_{0}^{\infty} \nu_{\mathrm{fin}} d \nu_{\mathrm{fin}} \\
    & \times \int_{0}^{\infty} \frac{n_{\mathrm{syn}}\left(\nu_{\mathrm{ini}},t\right)}{\nu_{\mathrm{ini}}}  f\left(q, \Gamma_{\epsilon}\right)  \theta(1-q)  \theta\left(q-1 / 4 \gamma^{2}\right) d v_{\mathrm{ini}},
\end{aligned}
\end{equation}
where $\nu_{\mathrm{ini}}$ and $\nu_{\mathrm{fin}}$ are the frequencies of initial the synchrotron radiation photons and that of scattered photons, $\Gamma_{\epsilon}=4 \gamma h \nu_{\text {ini }} /(m_{e} c^{2})$, $q=h \nu_{\text{fin}} /(\Gamma_{\epsilon}(\gamma m_{e} c^{2}-h \nu_{\text{fin }}))$, $f(q, \Gamma_{\epsilon})=2 q \ln q+(1+2 q)(1-q)+0.5(1-q)(\Gamma_{\epsilon} q)^{2} /(1+\Gamma_{\epsilon} q)$, and $\theta(x)$ is the step function. The seed synchrotron photon number density $n_{\mathrm{syn}}$ is  
\begin{equation}
    n_{\mathrm{syn}}(\nu, t)=\frac{L_{\nu, \mathrm{syn}}(t)}{4 \pi R_{\mathrm{n}}^{2}(t) c}  \frac{1}{h \nu}  \bar{U}, 
\end{equation}
where $L_{\nu, \mathrm{syn}}$ is the synchrotron radiation luminosity (see Section \ref{Lsyn}), and $\bar{U}\sim 2.24$ \citep{Atoyan1996} is used in our calculations. The adiabatic cooling is given by
\begin{equation}
    \dot{\gamma}_{\mathrm{ad}}(\gamma, t)=-\frac{1}{3} \gamma  \frac{d \ln V_{\mathrm{n}}}{d t}=-\frac{\dot{R}_{\mathrm{n}}}{R_{\mathrm{n}}} \gamma.
\end{equation}

\subsection{The synchrotron radiation of MWN}\label{Lsyn}
\begin{figure}
    \centering
    \includegraphics[width = 1\textwidth]{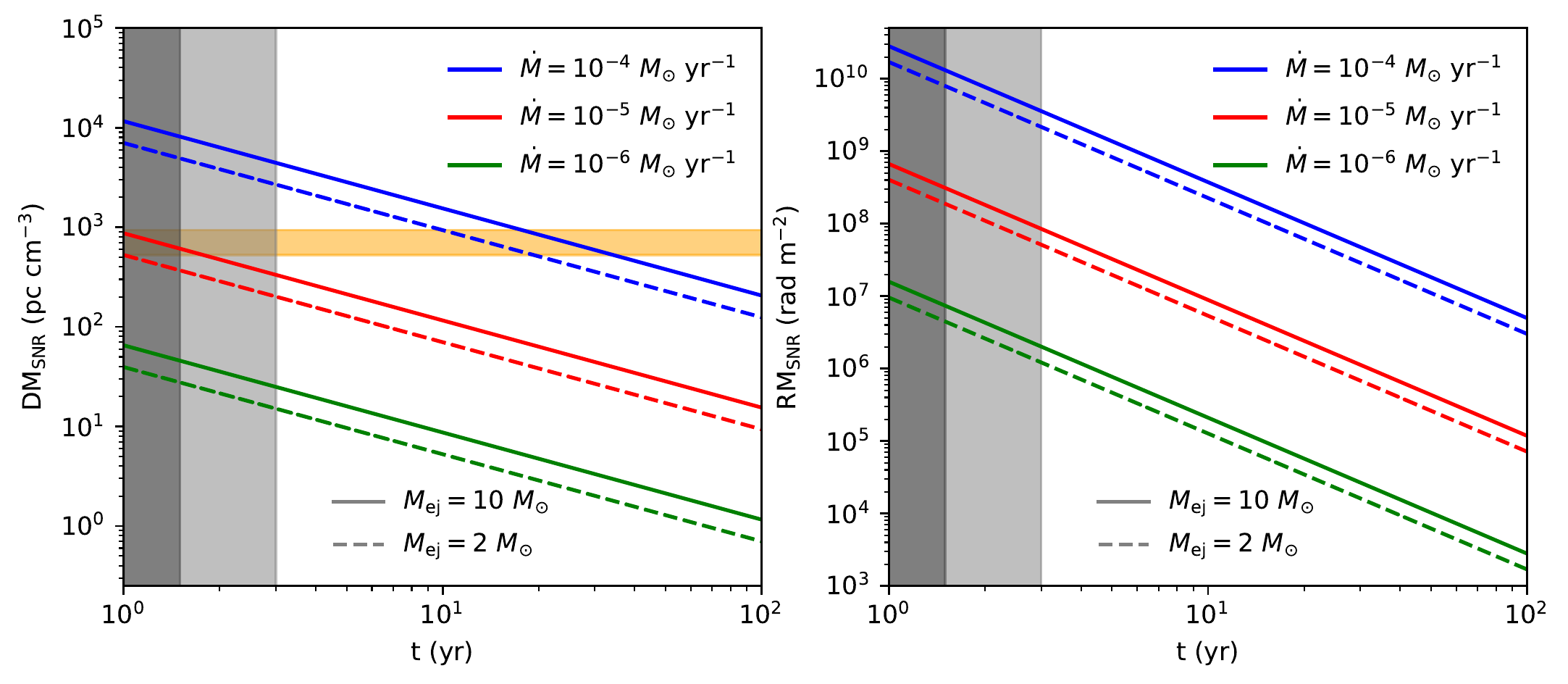}
    \caption{DM (left panel) and RM (right panel) contributed by SNR based on Equations (24) and (45)-(48) of \cite{Zhao2021}. We adopt the typical parameters of SNRs: the explosion energy $E_{\text{SN}}\sim1\times 10^{51}$ erg, the power-law index of outer ejecta $n=10$, ionization fractions of unshocked ejecta $\eta=0.1$, the wind velocity of progenitors $v_{\text{w}}=10$ km s$^{-1}$ and $\epsilon_{\mathrm{B}}=0.1$. The solid and dashed lines represent the case of $M_{\text{ej}}=10$ $M_{\odot}$ and $M_{\text{ej}}=2$ $M_{\odot}$, respectively. The blue, red and green lines represent different progenitors' mass-loss rates. The orange shading is the range of estimated $\mathrm{DM}_{\text {source }}$. We can see that only the case of $\dot{M}=10^{-4}$ $M_{\odot}$ yr$^{-1}$ and the source age $t_{\text{age}}=10-30$ yr can provide the large enough DM required by observations. The SNR will be optically thick for 3 yr and 1.5 yr for $M_{\text{ej}}=10$ $M_{\odot}$ and $M_{\text{ej}}=2$ $M_{\odot}$, which is shown in gray and black shaded regions.}
    \label{DMRM_SNR}
\end{figure}

The spectral power of synchrotron radiation is 
\begin{equation}
    P_{\nu}=\frac{2 e^{3} B_{\mathrm{n}}}{\sqrt{3} m_{\mathrm{e}} c^{2}} F\left(\frac{\nu}{\nu_{\mathrm{c}}}\right),
\end{equation}
where $\nu_{\mathrm{c}}=\gamma^{2} e B / 2 \pi m_{\mathrm{e}} c$ is the characteristic frequency, $F(x)=x \int_{x}^{+\infty} K_{5 / 3}(k) d k$ and $K_{5 / 3}(k)$ is the 5/3 order modified Bessel function. The emissivity and absorption coefficients of synchrotron radiation is 
\begin{equation}
    j_{\nu}=\int \frac{n_{e,\gamma} P_{\nu}(\gamma)}{4 \pi} d \gamma,
\end{equation}
\begin{equation}
    \alpha_{\nu}=-\int \frac{\gamma^{2} P_{\nu}(\gamma)}{8 \pi m_{\mathrm{e}} \nu^{2}} \frac{\partial}{\partial \gamma}\left(\frac{n_{e,\gamma}}{\gamma^{2}}\right) d \gamma.
\end{equation}
The synchrotron radiation luminosity considering the synchrotron self-absorption (SSA) is
\begin{equation}
    L_{\nu}=4 \pi^{2} R_{\mathrm{n}}^{2} \frac{j_{\nu}}{\alpha_{\nu}}\left(1-e^{-\alpha_{\nu} R_{\mathrm{n}}}\right).
\end{equation}
In addition to SSA, free-free absorption due to the ejecta is also important for radio signals from a young magnetar. From the study of DM and RM evolution of FRB 121102 \citep{Zhao2021}, the associated magnetar is in a clean environment, which means that the magnetar is born in the merger of two compact stars. For the merger channel, the ejecta mass is $\sim 0.001-0.1$ $M_{\odot}$, whose free-free absorption process is not obvious. However, for SNe channel, free-free absorption due to the ejecta can not be neglected. The free–free optical depth of the ejecta is \citep{WW2020,Zhao2021}
\begin{equation}
    \tau_{\mathrm{ff}, \mathrm{ej}}=\alpha_{\nu,\mathrm{ff}} \Delta R \simeq 2.06 \eta^{2} Y_{\mathrm{e}, 0.2}^{2} M_{\mathrm{ej},1}^{2} T_{\mathrm{ej}, 4}^{-3 / 2} \nu_{9}^{-2} v_{\mathrm{ej},9}^{-5} t_{\mathrm{yr}}^{-5},
\end{equation}
where $\eta$ is the ionization fraction, $Y_{\mathrm{e},0.2}=Y_{\mathrm{e}}/0.2$ is the the electron fraction, $M_{\mathrm{ej},1}=M_{\mathrm{ej}}/1 M_{\odot} $ is the ejecta mass, $T_{\mathrm{ej},4}=T_{\mathrm{ej}}/10^{4}$K is the ejecta temperature. Due to the free-free absorption of electrons, the SNR will be optically thick for 3 yr and 1.5 yr for $M_{\text{ej}}=10$ $M_{\odot}$ and $M_{\text{ej}}=2$ $M_{\odot}$, which is shown in gray and black shaded regions in Figures \ref{sp} and \ref{DMRM_SNR}, respectively.

\section{Numerical results}\label{result}
From the dynamics equations of MWN, we know that the nebula/ejecta velocity is mainly accelerated by the rotational energy injection and is almost constant for the time of our interest. The assumption of $v_{\text{n}}\simeq v_{\text{ej,f}}$ is a good approximation \citep{Metzger2014,Kashiyama2016} for $R_{\text{n}}>R_{\text{ej}}$. In our calculations, we take $v_{\text{ej,f}}=0.1 c$, which is the mean ejecta velocity of SN Ib/Ic \citep{Soderberg2012} and compact binary mergers.
Following \cite{Margalit2018}, $t_0=0.2$ yr and $\alpha=1.3$ is used in this work. The energy fraction $\epsilon_{\mathrm{B}}=0.1$ and $\epsilon_{\mathrm{B}}+\epsilon_{\mathrm{e}}\sim 1$ is used. The injection spectrum index $p_1=1.3$ and $p_2=2.5$ is taken from \cite{Law2019,Mondal2020}. The exact value of $\gamma_{\min}$ and $\gamma_{\max}$ is not important as long as its value is small or large enough. 
The main parameters are the interior magnetic field $B_{\text{int}}$, the source age $t_{\text{age}}$ and the break Lorentz factor $\gamma_{\text{b}}$.

\begin{figure}
    \centering
    \includegraphics[width = 1\textwidth]{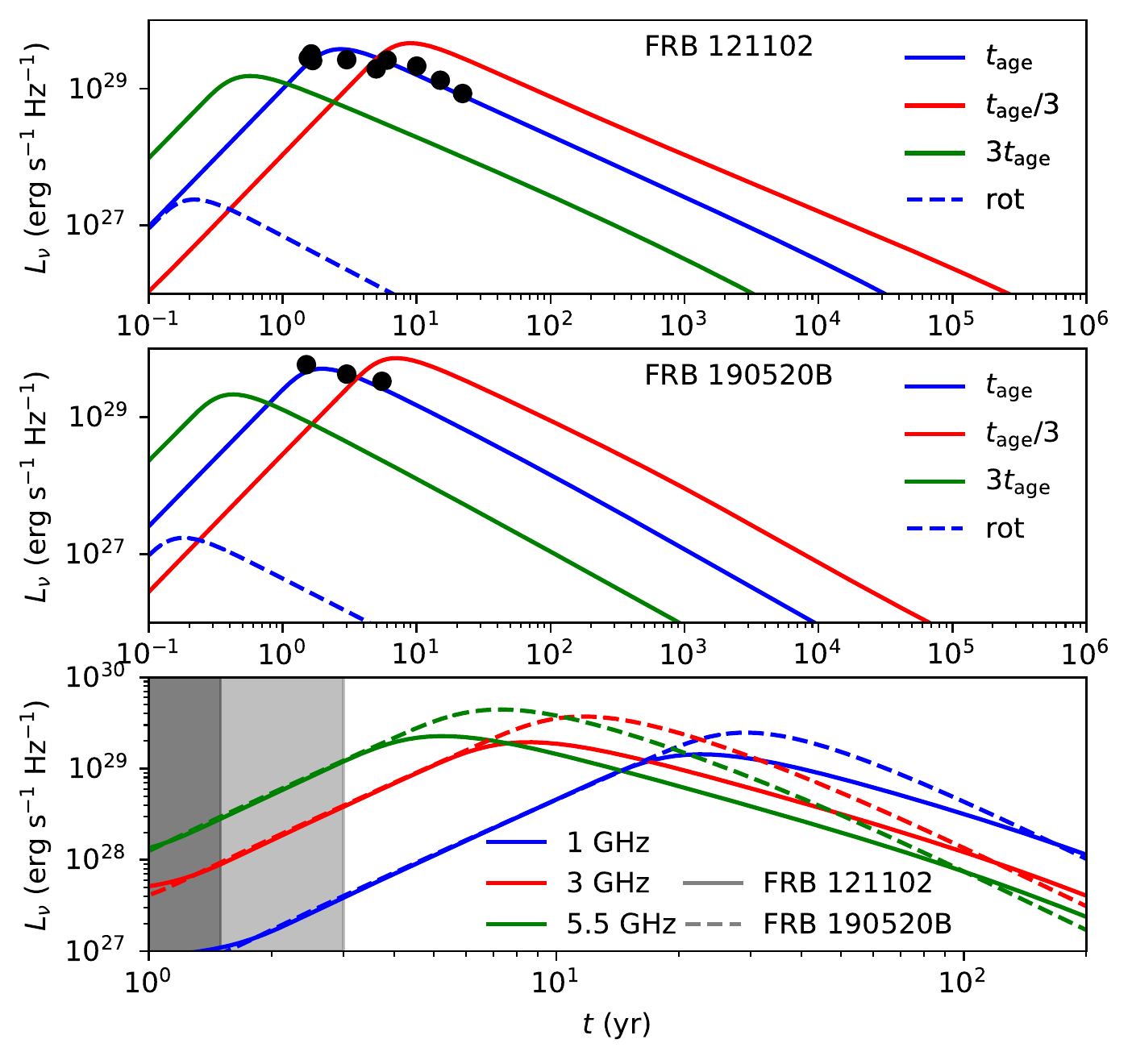}
    \caption{The spectral energy distribution for the PRS associated with FRB 121102 (top panel) and FRB 190520B (middle panel). The main parameters are $B_{\text{int}}=2.76\times 10^{16}$ G, $t_{\text{age}}=14$ yr and $\gamma_{\text{b}}=5\times10^{4}$ for FRB 121102 , and $B_{\text{int}}=3.41\times 10^{16}$ G, $t_{\text{age}}=22$ yr and $\gamma_{\text{b}}=5\times10^{3}$ for FRB 190520B. The red, blue and green solid lines represent the spectra observed epoch at $t=t_{\text{age}}/3$, $t=t_{\text{age}}$ and $t=3t_{\text{age}}$. The case of rotational energy injection is also plotted in dashed line at $t=t_{\text{age}}$ for comparison, whose dipole magnetic field is estimated as $B_{\text{dip}}=0.1B_{\text{int}}$ and initial spin period $P_{i}=5$ ms is taken. The black circles are the observations values from \cite{Chatterjee2017} for FRB 121102 and \cite{Niu2021} for FRB 190520B. The light curves of PRS (bottom panel) at 1 GHz, 3 GHz, and 5.5 GHz (blue, red, and green lines, respectively) for FRB 121102 and FRB 190520B (solid and dashed lines, respectively). The SNR will be optically thick for 3 yr and 1.5 yr for $M_{\text{ej}}=10$ $M_{\odot}$ and $M_{\text{ej}}=2$ $M_{\odot}$, which is shown in gray and black shaded regions.}
    \label{sp}
\end{figure}

The spectral energy distribution is shown in Figure \ref{sp}. The electron density $n_e$ and the nebula magnetic field $B_{\text{n}}$ are from the solutions of Equations (\ref{ne}) and (\ref{Bn}). We find that parameters $B_{\text{int}}=2.76\times 10^{16}$ G, $t_{\text{age}}=14$ yr and $\gamma_{\text{b}}=5\times10^{4}$ can reproduce the spectrum of the PRS associated with FRB 121102 \citep{Chatterjee2017}, and $B_{\text{int}}=3.41\times 10^{16}$ G, $t_{\text{age}}=22$ yr and $\gamma_{\text{b}}=5\times10^{3}$ for that of FRB 190520B \citep{Niu2021}. The source age of FRB 121102 we guessed as is to be roughly consistent with previous study \citep{Yang2019,Margalit2018,Zhao2021}. For FRB 190520B, the source age is given by the estimated from the DM evolution (see Section \ref{DM}). The red, blue and green solid lines represent the observed epoch at $t=t_{\text{age}}/3$, $t=t_{\text{age}}$ and $t=3t_{\text{age}}$. The case of rotational energy injection is also plotted as dashed line at $t=t_{\text{age}}$ for comparison, whose dipole magnetic field is estimated under the assumption of $B_{\text{dip}}=0.1B_{\text{int}}$ \citep{Levin2020} and initial spin period $P_{i}=5$ ms is taken. The light-curves at 1 GHz, 3 GHz, and 5.5 GHz (blue, red, and green lines, respectively) for FRB 121102 and FRB 190520B (solid and dashed lines, respectively) are shown in the bottom panel in Figure \ref{sp}. The size of MWN we obtained is $\sim 0.4$ pc for FRB 121102, which satisfies the constraints $<0.7$ pc given by very long baseline interferometry (VLBI, \citealt{Marcote2017}).

The DM and RM from the relativistic electrons in MWN are
\begin{eqnarray}
	\mathrm{DM_{MWN}} =R_{\mathrm{n}} \cdot \int \frac{n_{e,\gamma}}{\gamma^{2}} d \gamma, \\
	\mathrm{RM_{MWN}} =\frac{e^{3}}{2 \pi m_{e}^{2} c^{4}} R_{\mathrm{n}}  B_{\mathrm{n}} \cdot \int \frac{n_{e,\gamma}}{\gamma^{2}} d \gamma,
\end{eqnarray}
where the electron density $n_e$ and the nebula magnetic field $B_{\text{n}}$ are from the solutions of Equations (\ref{ne}) and (\ref{Bn}). We obtain that the DM from the MWN is $<1-10$ pc cm$^{-3}$ and RM is $<10^4-10^5$ rad m$^{-2}$. The contributions from the MWN are negligible compared with the SNR (see Section \ref{DM}). 

\section{Long-term DM evolution}\label{DM}
FRB 190520B has been reported in a dense environment, and the estimated $\mathrm{DM}_{\text {host}}\simeq 902^{+88}_{-128}$ pc cm$^{-3}$ \citep{Niu2021} is nearly ten times higher than other FRB host galaxies. The DM of FRB 190520B systemically decreases with the rate  $-0.09 \pm 0.02$ pc cm$^{-3}$ day$^{-1}$ together with some irregular variations \citep{Niu2021}.  In our model, the long-term DM variation is from the expanding SNR \citep{Yang2017,Piro2018,Zhao2021,Katz2021} and the random variations may be caused by turbulent motions of filament \citep{Katz2021b}.

\subsection{The DM from the local environment}
For cosmological FRBs, the observed DM or RM contains the contributions of the Milky Way (MW), the Milky Way halo, the intergalactic medium (IGM), the host galaxy and the local environment of FRBs:
\begin{equation}
    \mathrm{DM}_{\mathrm{obs}}=\mathrm{DM}_{\mathrm{MW}}+\mathrm{DM}_{\text {halo }}+\mathrm{DM}_{\mathrm{IGM}}+\frac{\mathrm{DM}_{\text {host}}+\mathrm{DM}_{\text {source }}}{1+\mathrm{z}},
\end{equation}
\begin{equation}
    \mathrm{RM}_{\mathrm{obs}}=\mathrm{RM}_{\mathrm{MW}}+\mathrm{RM}_{\text {halo }}+\mathrm{RM}_{\mathrm{IGM}}+\frac{\mathrm{RM}_{\text {host }}+\mathrm{RM}_{\text {source }}}{(1+z)^{2}}.
\end{equation}
Using the IllustrisTNG simulation, \cite{Zhang2020} found that the DM contributed by FRB 190520B-like host galaxies at $z\sim0.2$ is $\sim50-250$ pc cm$^{-3}$. Therefore, the DM from the source of FRB 190520B can be inferred to be $\sim 524-940$ pc cm$^{-3}$. Following our previous work \citep{Zhao2021}, the DM from the local environment of FRBs is given by
\begin{equation}
    \mathrm{DM}_{\text {source}}=\mathrm{DM}_{\mathrm{MWN}}+\mathrm{DM}_{\mathrm{unsh}, \mathrm{ej}}+\mathrm{DM}_{\mathrm{sh}, \mathrm{ej}}+\mathrm{DM}_{\mathrm{sh}, \mathrm{ISM}}+\mathrm{DM}_{\mathrm{unsh}, \mathrm{ISM}},
\end{equation}
where $\mathrm{DM}_{\mathrm{MWN}}$, $\mathrm{DM}_{\mathrm{unsh}, \mathrm{ej}}$, $\mathrm{DM}_{\mathrm{sh}, \mathrm{ej}}$, $\mathrm{DM}_{\mathrm{sh}, \mathrm{ISM}}$ and $\mathrm{DM}_{\mathrm{unsh}, \mathrm{ISM}}$ are the contributions from the MWN, unshocked ejecta, shocked ejecta, shocked ISM and unshocked ISM, respectively. Usually, $\mathrm{DM}_{\mathrm{unsh}, \mathrm{ISM}}$ is negligible because of the low ionization fraction. The unshocked region is not magnetized, so the RM from the source only contributed by three parts
\begin{equation}
    \mathrm{RM}_{\text {source}}=\mathrm{RM}_{\mathrm{MWN}}+\mathrm{RM}_{\mathrm{sh}, \mathrm{ej}}+\mathrm{RM}_{\mathrm{sh}, \mathrm{ISM}}.
\end{equation}
The contributions from the MWN are also negligible compared with the SNR. The total DMs and RMs from the SNR are shown in Figure \ref{DMRM_SNR}. Our calculations are based on Equations (24) and (45)-(48) of \cite{Zhao2021}. We adopt the typical parameters of SNRs: the explosion energy $\mathcal{E_{\text{SN}}}\sim1\times 10^{51}$ erg, the power-law index of outer ejecta $n=10$, ionization fractions of unshocked ejecta $\eta=0.1$, the wind velocity of progenitors $v_{\text{w}}=10$ km s$^{-1}$ and $\epsilon_{\mathrm{B}}=0.1$. The solid and dashed lines represent the case of $M_{\text{ej}}=10$ $M_{\odot}$ and $M_{\text{ej}}=2$ $M_{\odot}$, respectively. The blue, red and green lines represent different progenitors' mass-loss rate. The orange shading is the range of estimated $\mathrm{DM}_{\text {source }}$. We can see that only the case of $\dot{M}=10^{-4}$ $M_{\odot}$ yr$^{-1}$ and the source age $t_{\text{age}}=10-30$ yr can provide the large enough DM. The constraint on the source age is consistent with that derived from PRS in the previous section. 

\subsection{Fitting Results}
We assume that the variations of DM is only from $\mathrm{DM}_{\text {source}}$. Thus, we can define the unchangeable $\mathrm{DM}_{\text {other}}=\mathrm{DM}_{\text {obs}}-\mathrm{DM}_{\text {source}} $ for facilitate fitting. From the estimation of $\mathrm{DM}_{\text {source}}$ above, we can get $\mathrm{DM}_{\text {other}}\sim 270-686$ pc cm$^{-3}$. The Markov Chain Monte Carlo (MCMC) method performed by Python
package \texttt{emcee}\footnote{\url{emcee.readthedocs.io}} \citep{Foreman-Mackey2013} is used to estimate the parameters $\mathrm{DM}_{\text {other}}$ and the age of the source $t_{\text{age}}$. The $\chi^2$ for the observed DMs is
\begin{equation}
    \chi_{\mathrm{DM}}^{2}=\sum_{i=1}^{n} \frac{\left(\mathrm{DM}_{\mathrm{SNR},i}-\mathrm{DM}_{\mathrm{obs},i}\right)^{2}}{\sigma^2},
\end{equation}
where $\mathrm{DM}_{\mathrm{SNR}}$ is the DM from SNR given by our model, and $\mathrm{DM}_{\mathrm{obs}}$ and $\sigma$ is the observed DM and uncertainties in the frame of observers \citep{Niu2021}. The likelihood is 
\begin{equation}
    \mathcal{L} \propto \exp \left[-\left(\chi_{\mathrm{DM}}^{2}\right) / 2\right].
\end{equation}

\begin{figure}
    \centering
    \includegraphics[width = 1\textwidth]{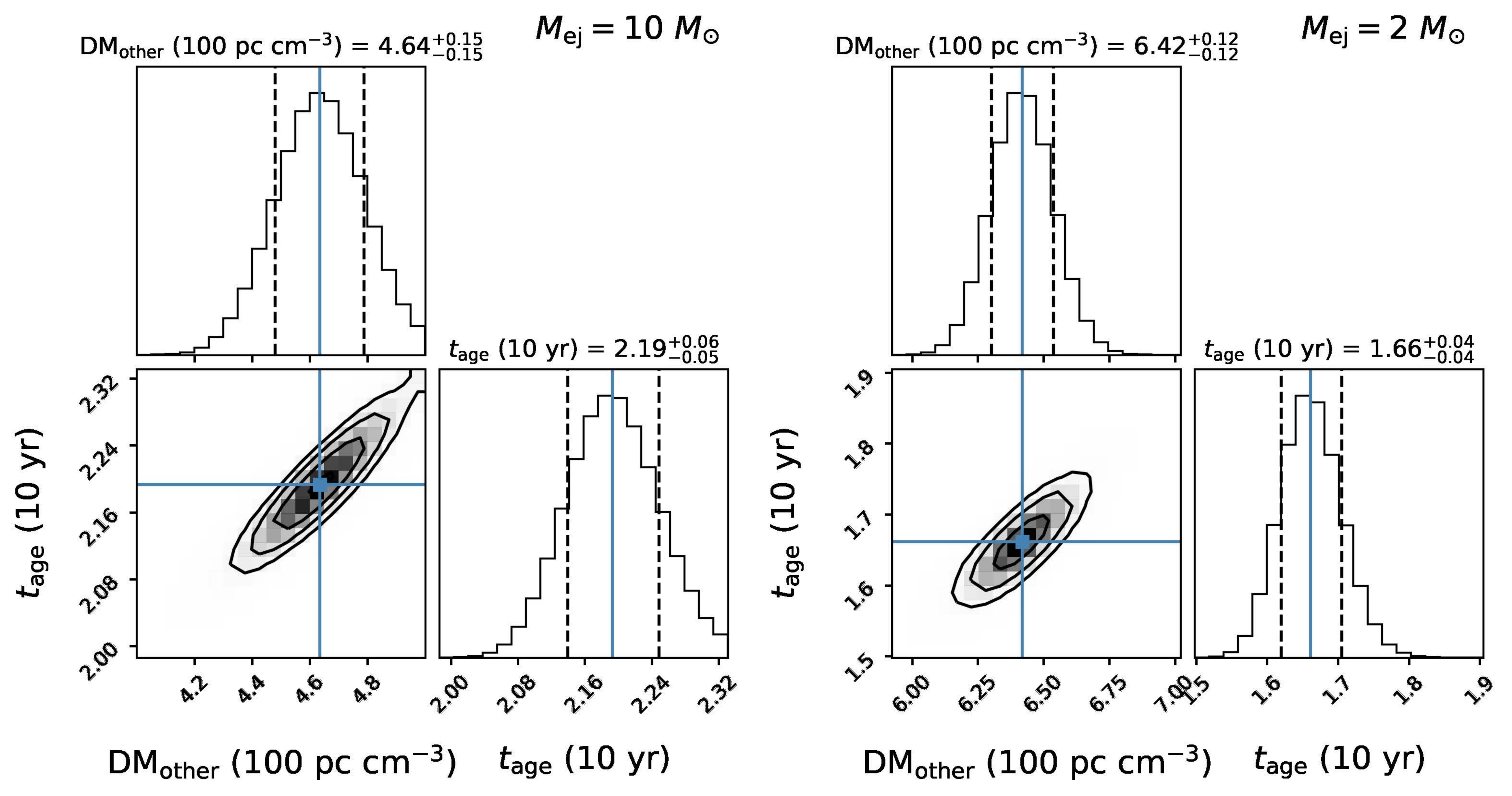}
    \caption{Posteriors parameters of DM model fit to FRB 190520B for the case of $M_{\text{ej}}=10$ $M_{\odot}$ (left panel) and $M_{\text{ej}}=2$ $M_{\odot}$ (right panel) performed by MCMC method. The posterior probability of the parameters of $\mathrm{DM}_{\text {other}}=\mathrm{DM}_{\text {obs}}-\mathrm{DM}_{\text {source}} $ and $t_{\text{age}}$ are plotted in the histograms, and the 1-$\sigma$ range is shown in the dashed vertical lines. The medians are shown in blue lines. The contours indicate the parameter space with 1-$\sigma$, 2-$\sigma$ and 3-$\sigma$ range. If we assume that the typical SN ejecta mass is $2M_{\odot}< M_{\text{ej}}<10M_{\odot}$, the source age of FRB 190520B can be estimated as $16-22$ yr. In the same way, we have $\mathrm{DM}_{\text {other}}=449-654$ pc cm$^{-3}$ ($\mathrm{DM}_{\text {SNR}}=556-761$ pc cm$^{-3}$).}
    \label{corner}
\end{figure}

\begin{figure}
    \centering
    \includegraphics[width = 1\textwidth]{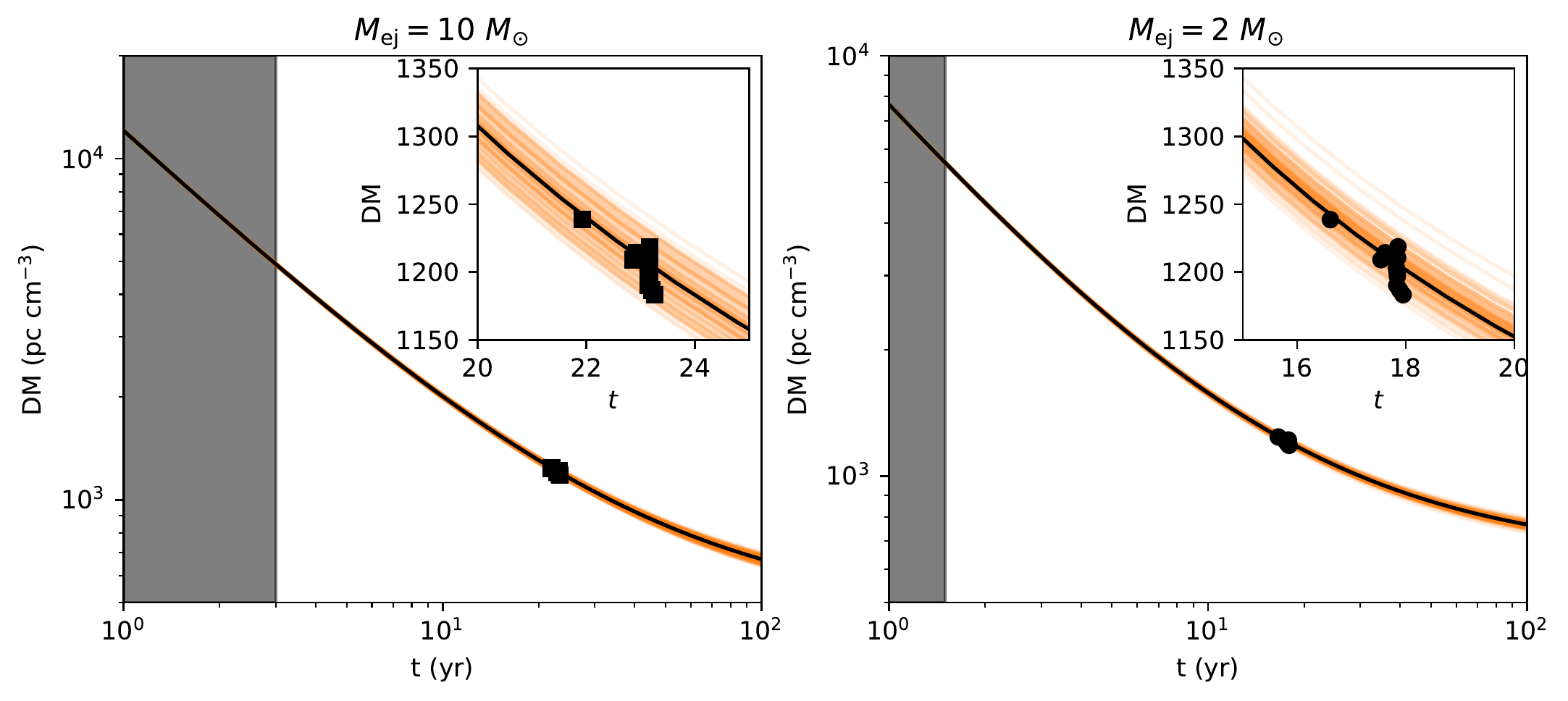}
    \caption{Samples of DM$_{\text{SNR}}$ for $M_{\text{ej}}=10$ $M_{\odot}$ (left panel) and $M_{\text{ej}}=2$ $M_{\odot}$ (right panel) from the MCMC method (orange curves). Black squres and circles are the DM$_{\text{obs}}$ of FRB 190520B from \cite{Niu2021}. Black lines represent the best-fit values given by the MCMC method. The shaded regions represent the SNR is opaque to radio signals of $\nu \sim$ 1 GHz. }
    \label{DM_SNR}
\end{figure}

The posterior corner plots obtained from fitting the models of two typical ejecta mass models ($M_{\text{ej}}=10$ $M_{\odot}$ and $M_{\text{ej}}=2$ $M_{\odot}$) to the deta are shown in Figure \ref{corner}. Our best-fit parameters are shown in blue solid lines, and the parameters with 1-$\sigma$ ranges are shown in dashed lines. For $M_{\text{ej}}=10$ $M_{\odot}$, we find $\mathrm{DM}_{\text {other}}=464\pm 15$ pc cm$^{-3}$ ($\mathrm{DM}_{\text {SNR}}= 746\pm 15$ pc cm$^{-3}$) and the source age is $21.9\pm 0.5 $ yr. For $M_{\text{ej}}=2$ $M_{\odot}$, we find $\mathrm{DM}_{\text {other}}=642\pm 12$ pc cm$^{-3}$ ($\mathrm{DM}_{\text {SNR}}=568\pm 12$ pc cm$^{-3}$) and the source age is $16.6\pm 0.4 $ yr. The DM evolution after the SN explosion is plotted in Figure \ref{DM_SNR}. Due to the free-free absorption of electrons, the SNR will be optically thick for 3 yr and 1.5 yr for $M_{\text{ej}}=10$ $M_{\odot}$ and $M_{\text{ej}}=2$ $M_{\odot}$, respectively. The shaded regions represent the SNR is opaque to radio signals of $\nu \sim$ 1 GHz. If we assume that the typical SN ejecta mass is $2M_{\odot}< M_{\text{ej}}<10M_{\odot}$, the source age of FRB 190520B can be estimated as $16-22$ yr. In the same way, we have $\mathrm{DM}_{\text {other}}=449-654$ pc cm$^{-3}$ ($\mathrm{DM}_{\text {SNR}}=556-761$ pc cm$^{-3}$).

When DM$_{\text{source}}$ is taken into consideration, the DM$_{\text{host}}$ of FRB 190520B is not special, which is consistent with that derived from the IllustrisTNG simulation \citep{Zhang2020}. The DM decline will continue for another few decades, and then DM will trend to be stable when $\mathrm{DM}_{\text {other}}\gg\mathrm{DM}_{\text {SNR}}$. Finally, we have $\mathrm{DM}\sim\mathrm{DM}_{\text {other}}$ for $t\sim100-1000$ yr. The unchangeable DM has been reported for FRBs (e.g., FRB 180916, see \citealt{2020Natur.582..351C,Pastor-Marazuela2021,Nimmo2021}). For FRB 121102, the estimated RM$_{\mathrm{MWN}}\sim10^4-10^5$ rad m$^{-2}$ at $t_{\text{age}}=14$ yr is consistent with the study of DM and RM evolution (\citealt{Zhao2021}, their estimated age starts on 2012). For FRB 190520B, the large RM is from the young SNR (RM $\sim10^{7}-10^{8} \mathrm{rad} ~\mathrm{m}^{-2}$).

\section{Summary}\label{conclusion}
In this letter, we argue that the magnetar associated with FRB 190520B is embedded in the `composite' of MWN and SNR. Due to the energy injection of the young magnetar (16-22 yr), the wind nebula and the SN ejecta will evolve together. The observed PRS is from the synchrotron radiation of the nebula. The dense SNR ejecta and the shocked shell contribute the observed DM and RM. Our model can simultaneously explain the luminous PRS, decreasing DM and extreme RM of FRB 190520B. Our conclusions are summarized as follows:
\begin{itemize}
    \item The compact PRSs associated with FRBs are from the synchrotron radiation of the MWNe. From the observed luminosities and spectra, we find the interior magnetic field $B_{\text{int}}= 2.76\times 10^{16}$ G, the source age $t_{\text{age}}= 14$ yr for FRB 121102, and $B_{\text{int}}= 3.41\times 10^{16}$ G, $t_{\text{age}}= 22$ yr for FRB 190520B.
    \item FRB 190520B is embedded in a dense SNR whose DM contribution is $\sim 746\pm 15$ pc cm$^{-3}$ and $\sim 568\pm 12$ pc cm$^{-3}$ for $M_{\text{ej}}=10$ $M_{\odot}$ and $M_{\text{ej}}=2$ $M_{\odot}$, respectively. Considering the DM from the SNR, the DM from the interstellar medium of FRB 190520B host galaxy is not special any more, which is consistent with that derived form the state-of-the-art IllustrisTNG simulation \citep{Zhang2020}. The DM decay rate $-0.09 \pm 0.02$ pc cm$^{-3}$ d$^{-1}$ can be well understood in the context of a SNR with the age of $16-22$ yr, well in the range required by the PRS. The decline will continue for another few decades, and then DM will trend to be stable. 
    \item For FRB 190520B, the large RM is from the young SNR (RM $\sim10^{7}-10^{8} ~\mathrm{rad} ~\mathrm{m}^{-2}$). The RM attributed to MWN is $<10^4-10^5$ rad m$^{-2}$ in our model, which is much lower than the lower limit given by \cite{Niu2021} and the contributions from SNR.
\end{itemize}

\section*{acknowledgements}
We thank the anonymous referee for helpful comments. We acknowledge Yuan-Pei Yang, Ling-Jun Wang, Fan Xu, Long Li, Abudushataer Kuerban, Younes George and Kohta Murase for helpful discussions, Jin-Jun Geng, Zhao Zhang and Qiao-Chu Li for the help of numerical calculations. This work was supported by the National Natural Science Foundation of China (grant No. U1831207), and the Fundamental Research Funds for the Central Universities
(No. 0201-14380045).

\bibliographystyle{aasjournal}
\bibliography{ms}

\end{document}